\documentstyle[epsf,graphicx]{mn}

\def\x2{$\chi^{2}$}

\def\asca{{\it ASCA }}
\def\rosat{{\it ROSAT }}

\def\x2{$\chi^{2}$}
\def\lunits{$\rm{erg\,s^{-1}}$}
\def\funits{$\rm{erg\,s^{-1}\,cm^{-2}}$}
\def\cunits{$\rm{cm^{-2}}$}

\newbox\grsign \setbox\grsign=\hbox{$>$} \newdimen\grdimen \grdimen=\ht\grsign
\newbox\simlessbox \newbox\simgreatbox \newbox\simpropbox
\setbox\simgreatbox=\hbox{\raise.5ex\hbox{$>$}\llap
     {\lower.5ex\hbox{$\sim$}}}\ht1=\grdimen\dp1=0pt
\setbox\simlessbox=\hbox{\raise.5ex\hbox{$<$}\llap
     {\lower.5ex\hbox{$\sim$}}}\ht2=\grdimen\dp2=0pt
\setbox\simpropbox=\hbox{\raise.5ex\hbox{$\propto$}\llap
     {\lower.5ex\hbox{$\sim$}}}\ht2=\grdimen\dp2=0pt

\begin{document}

\title[X-ray observations of IRAS00317-2142] {X--ray observations of the
'composite'  Seyfert/star-forming galaxy IRAS00317-2142}

\author[I. Georgantopoulos]
       {I. Georgantopoulos  \\
  Astronomical Institute, National Observatory of Athens, Lofos Koufou, 
 Palaia Penteli, 15236, Athens, Greece } 

\maketitle

\label{firstpage}

\begin{abstract}

I present \asca  observations of IRAS00317-2142,
 the most luminous ($L_x\sim 10^{43}$ \lunits 0.1-2 keV)  
 of the 'composite' 
 class of galaxies. This enigmatic  class of objects
 presents narrow-emission line optical spectra  
 classifying these galaxies as star-forming on the 
 basis of the diagnostic emission line ratios; yet, 
  the presence of weak $H_{\alpha}$ broad wings also 
 suggests the presence of a weak or obscured AGN.   
 The \asca spectrum can be represented with a power-law 
 with a photon index of $\Gamma=1.76\pm 0.08$.  
 Strong variability is detected (by about a factor 
 of three) between the \rosat 
 and \asca 1-2 keV flux. These characteristics 
 clearly suggest an AGN origin for the X-ray emission. 
 However, the precise nature of this AGN remains still 
 uncertain. 
 There is no  evidence for a high absorbing column density 
  above the Galactic. Moreover,  
 there is no strong evidence for an Fe line at 6.4 keV, 
 with the 90 per cent upper limit on the equivalent width being 0.9 keV.
 Thus the X-ray spectrum  is consistent with an
 unobscured  Seyfert-1 interpretation. 
 This discrepancy with the optical spectrum, 
 may be explained by either    
 a strong  star-forming component or 
 a 'dusty' ionised absorber.  
   Finally, the possibility 
 that IRAS00317-2142 may harbour a heavily obscured AGN 
 where the X-ray emission is mainly due to scattered light,
  appears less plausible due to the high value of the 
 $f_{\rm x}/f_{\rm [OIII]}$ ratio which is more 
 indicative of unobscured Seyfert-1 type AGN.  
  \end{abstract}

\begin{keywords}
galaxies: starburst -- galaxies: AGN - galaxies:individual: 
IRAS00317-2142 -- X-rays: AGN
\end{keywords}

\section{INTRODUCTION}

The \rosat All-Sky Survey, RASS, 
 (Voges et al. 1996) provided a wealth of information on 
the X-ray properties of nearby AGN. 
 The cross-correlation of the RASS with the IRAS Point Source Catalogue
 (Boller et al. 1995, 1998),
 revealed a new interesting class of objects, named 'composites' 
 by Moran et al. (1996), with  high luminosities $L_x\sim 10^{42-43}$ \lunits. 
 Their main characteristic is that their [OIII] 
 lines are broader than all other narrow lines,
 forbidden or permitted (Moran et al. 1996). 
  The diagnostic emission line ratio diagrams (Veilleux \& Osterbrock 
 1987)  classify these objects as  star-forming galaxies. 
 Yet, some of the 'composites' present broad $H_\alpha$ wings 
  suggesting the presence of  a weak 
 or obscured AGN. The origin of the powerful X-ray emission 
 remains unknown in these objects. Interestingly, 
  the 'composites'  show very similar 
  optical spectra with those of the  
 narrow-line galaxies  (NLXGs) detected in deep 
 \rosat surveys (eg Boyle et al. 1995). 
The NLXGs have narrow line optical spectra,
 some with weak broad wings. Yet, the  powerful 
 X-ray emission $L_x\sim 10^{42-43}$ \lunits 
 strongly suggests the presence 
 of a relatively weak or a heavily obscured AGN. 
 
The most luminous object in the 'composite' sample,
 IRAS00317-2142 has a luminosity of $L_x\sim 10^{43}$ \lunits (0.1-2 keV) 
 at a redshift of z=0.0268 (Moran et al. 1996). 
 This galaxy belongs to a small group 
 of galaxies, Hickson 4 (Ebeling et al. 1994). Its optical spectrum 
 has been discussed in detail by Coziol et al. (1993).  
 The diagnostic emission line ratios  
 (eg $H_\alpha/[NII]$ vs $H_\beta/[OIII]$) 
 classify it as an HII galaxy. Nevertheless, 
 its high X-ray luminosity   
 as well as the presence of  a faint $H_\alpha$ wind 
 in its optical spectrum (Coziol et al. 1993) would clearly 
 assign an AGN classification to this object. 
 Here, I present an analysis of the X-ray properties 
 of IRAS00317-2142  using data from both the \asca and the \rosat 
 missions. The goals are to understand the 
 origin of the X-ray emission  in this enigmatic class of objects,
  its  possible 
 relation to the NLXGs present in the deep \rosat surveys 
   and  finally to constrain the 'composite' galaxy 
   contribution to the X-ray background.  

\section{OBSERVATIONS AND DATA REDUCTION}

IRAS00317-2142 was observed with \asca (Tanaka, Inoue \& Holt 1994)
 between the  11 and 12th 
 of December 1995. I have used the ``Rev2'' processed data 
 from the HEASARC database at the Goddard Space 
Flight Center. For the selection criteria applied on Rev2 data, 
 see the \asca Data ABC guide (Yaqoob 1997). 
 Data reduction was performed using FTOOLS v4.2.
 The net exposure time is about 40 ksec and 37 ksec for 
 the GIS and the SIS detectors respectively.
The two GIS and the two SIS detectors on-board ASCA have an 
 energy range roughly between 0.8-10 keV and 0.5-10 keV respectively.
The energy resolution of the SIS CCD detectors is 2 per cent at 6 keV 
while of the GIS detectors is 8 per cent at the same energy. 
For more details on the \asca detectors see Tanaka et al. (1994).  
A circular extraction cell for the source of 2 arcminute 
radius has been used.
Background counts were estimated from source-free regions 
 on the same images.    
The observed flux in the 2-10keV band is
$f_{2-10keV}\simeq 8\times10^{-13}$
\funits  while the one in the 1-2 keV band is 
$f_{1-2keV}=2.5\times10^{-13}$\funits; the  fluxes are 
 estimated using the best-fit power law model below ($\Gamma=1.8$). 
 
\rosat observed IRAS00317-2142 on two occasions. 
 It was first detected during the RASS (exposure time 340 s).
 Its RASS flux is $2.7\pm0.46\times 10^{-12}$ (0.1-2 keV),
 (Moran et al.  1996). 
 It was also observed by \rosat  PSPC as a target
 during a pointed observation between the 22 and 23rd 
 of June 1992 (exposure time 9.3 ksec).  
 The derived flux was $2.7\pm0.05 \times 10^{-12}$ \funits 
 in the 0.1-2 band,
   in excellent agreement with the RASS flux.
 In the 1-2 keV band the flux is $ 7.3\times 10^{-13}$ \funits
 a factor of three above that of \asca  
 in the same band.  There is no evidence for 
 extension in the pointed \rosat PSPC image 
 (eg Pildis, Bregman \& Evrard 1995)  
 suggesting that the bulk of the X-ray emission 
 originates in IRAS00317-2142 rather than  
 being diffuse emission from  hot intergalactic gas
 in the galaxy group. Here, I re-analyse the \rosat data  
  in order to make comparisons with the \asca spectral fits
 and to  perform joint fits with the \asca data. 

Throughout this paper I adopt $ \rm H_\circ=50 km s^{-1} Mpc^{-1}$
 and $q_o=0$. For the spectral fitting I use XSPEC v.10. 
 I bin the data so that there are at least 20 counts per bin 
 (source and background). 
 Quoted errors to the best-fitting 
 spectral parameters are 90 per cent
confidence regions for one parameter of interest.

\section{ SPECTRAL ANALYSIS}

\subsection{The ASCA spectral fits}
I first fit a single power-law to the \asca data. These are consistent 
with a zero hydrogen column density, $N_H$. 
 Hence, hereafter, I have fixed the column to
 the Galactic column density ($N_H\sim 1.5\times 10^{20}$ $\rm cm^{-2}$).
The results of the \asca spectral fits are given in Table 1. 
 Entries with no associated error bars were fixed to this value
 during the fit.  The power-law slope 
$\Gamma \approx 1.8$, is consistent with the canonical spectral index
  of AGN (eg Nandra \& Pounds 1994). Although this simple model 
 provides an acceptable fit, $\chi^2=118.2/114$ 
 degrees of freedom (dof), I have also added a Gaussian line component
 to the fit (the energy and the line width fixed at 6.4 keV and 0.01 keV 
respectively) as this is a common feature in AGN spectra. The additional
 component marginally improves the fit ($\Delta\chi^2=2.4$ for one
 additional parameter); 
this is statistically significant at only the 90 per cent confidence level.
 The 90 per cent upper limit for the equivalent width 
 is 0.9 keV.  
In Fig. 1 the \asca spectrum together with 
 the best fit power-law model are given; the data residuals    
from the model are also plotted. The data have been rebinned in the plot  
 for clarity. A Raymond-Smith (RS) thermal model results in 
 a worse fit ($\chi^2=128.0/113$). The temperature 
 derived (kT=5.8 keV) is reminiscent of nearby 
 normal galaxies.    
 Next, I fit an ionised warm absorber model (eg Brandt, Fabian \& 
 Pounds 1996) in addition to the 
 Galactic column density. Indeed warm absorbers are detected in more than 
 50 per cent of Seyfert 1s (Brandt et al. 1999). 
 The temperature of the absorber is fixed at $T=10^{5} \rm K$
 (Brandt et al. 1999). The best fit 'warm' column density 
  is $N_H\sim 10^{22}$ \cunits while the 
 ionisation parameter is practically unconstrained. However, 
 $\Delta\chi^2\approx 2.4 $ for two additional parameters 
 and thus the warm absorber model does not represent a
 statistically significant improvement.  
  
I also attempt to fit a more complicated model with both a power-law and
 a RS component. This is because the optical spectrum strongly
 suggests the presence of a strong star-forming component. 
 The spectral fit above yields $\chi^2=115.3/112$; 
 the  inclusion of the additional RS component is 
 not statistically significant ($<1\sigma$). I  
 derive a spectral index of $\Gamma=1.7^{+0.10}_{-0.10}$.
  The RS component has a temperature of kT$\sim$ 0.2 keV  
 lower but consistent with those of 
   the star-forming regions in nearby galaxies (eg Read \& Ponman 1997).
 The abundance remains practically unconstrained and thus it 
 was fixed to the solar value ($Z=1$).  
 The luminosity of the RS component is $\sim 5\times 10^{41}$ 
 $\rm erg ~s^{-1}$ or about 25 per cent of the 
 total luminosity in the 0.5-2 keV band. 
  Finally, a power-law and RS model is fit where the 
 obscuring column densities are different in the two components. 
  For example in many Seyfert-2, the power-law component 
 is heavily obscured while the star-forming component 
 is outside the obscuring screen and 
 is relatively unobscured. 
 However, both best fit columns are close to the Galactic 
 disfavouring the above scenario.   

\subsection{\asca and \rosat fits}

 Due to the low energy  coverage of the \rosat PSPC  (0.1-2 keV) and
 its high effective area, it is quite instructive to fit the \rosat
 data as well. Nevertheless, the energy resolution of the \rosat
 PSPC is very limited   
($\Delta E/ E \sim$ 50 per cent) at 1 keV. RS spectra have been fit to
 the \rosat data by Saracco \& Ciliegi  (1995) and Pildis et al. (1994). 
They derive a temperature of kT$\sim$ 1 keV. 
Instead, I fit a power-law spectrum which is the standard model 
 for AGN spectra at least in the small \rosat band. 
 I obtain a spectral index of 
$\Gamma=2.63^{+0.04}_{-0.16}$ much steeper than the \asca fit. 
The column density is $N_H=3.4^{+1.0}_{-0.4}\times 10^{20}$ $\rm cm^{-2}$
 higher than the Galactic column ($\chi^2=82.1/82$ dof).
Finally, for the sake of completeness, I perform joint fits to the 
\asca and the \rosat data.
 However, bear in mind that this joint analysis has 
to be viewed with great caution. Recently Iwasawa Nandra \& Fabian (1999) 
 made spectral fits on {\it simultaneous} \asca and \rosat 
data of NGC5548. They demonstrate that the power-law fits 
 may differ as much as 
$\Delta\Gamma\approx 0.4$ even in the common \rosat/\asca 0.5-2 keV band. 
 The reason for this large discrepancy may be related with uncertainties 
 in the calibration of both the \asca and \rosat detectors.  
 A single power-law fit to the combined \asca and \rosat data 
 of IRAS00317-2142 yields $\Gamma=2.00^{+0.07}_{-0.07}$,
 $N_H=1.9\pm 0.2\times 10^{20}$ $\rm cm^{-2}$ 
 ($\chi^2=227.2/197$ dof). The power-law normalization is allowed to 
 vary freely between the \rosat and the \asca observation epoch. 
 Next, a RS component is added to the model. The 
 temperature is constrained to have an upper 
 limit of 1 keV, otherwise the resulting temperature 
 becomes unrealistically high ($\sim$ 15 keV). 
    The best fit temperature is $\rm kT=0.07^{+0.02}_{-0.01}$ keV 
  while the power-law slope is $\Gamma=1.9^{+0.12}_{-0.03}$. 
 Despite the inclusion of the additional RS  
 component the  fit did not improve ($\chi^2= 231.4/195$ dof).

\begin{table*}
\caption{The spectral fits results on the \asca data}
\begin{tabular}{cccccccc}
\hline
Model & $N_H${\small ($10^{20} \rm cm^{-2}$)} & $N_H^{warm}$ 
 {\small ($10^{20} \rm cm^{-2}$)} & $\Gamma$ & $Z$ & kT 
 {\small (keV)} &  $\xi$ &   $\chi^2 / dof$ \\ 
\hline 
power-law & 1.5 & - & $1.76^{+0.08}_{-0.08}$ & - & - & - &  118.2/114 \\
warm-absorber & 1.5 & $180^{+450}_{-170}$ & $1.82^{+0.10}_{-0.10}$ &  1 & -& 
  $358^{+\infty}_{-330}$ & 115.8/112 \\
Raymond-Smith (RS) & 1.5 & - &- & $0.56^{+0.48}_{-0.39}$ &  
 $5.8^{+1.5}_{-1.3}$ & - & 128.0/113 \\
power-law + RS  & 1.5 & - &$1.71^{+0.10}_{-0.10}$ &  1 & $0.18^{+0.47}_{-0.08}$
  & - & 115.3/112 \\
\hline
\end{tabular}
\end{table*}

\begin{figure*}
\rotatebox{270}{\includegraphics[height=11.5cm]{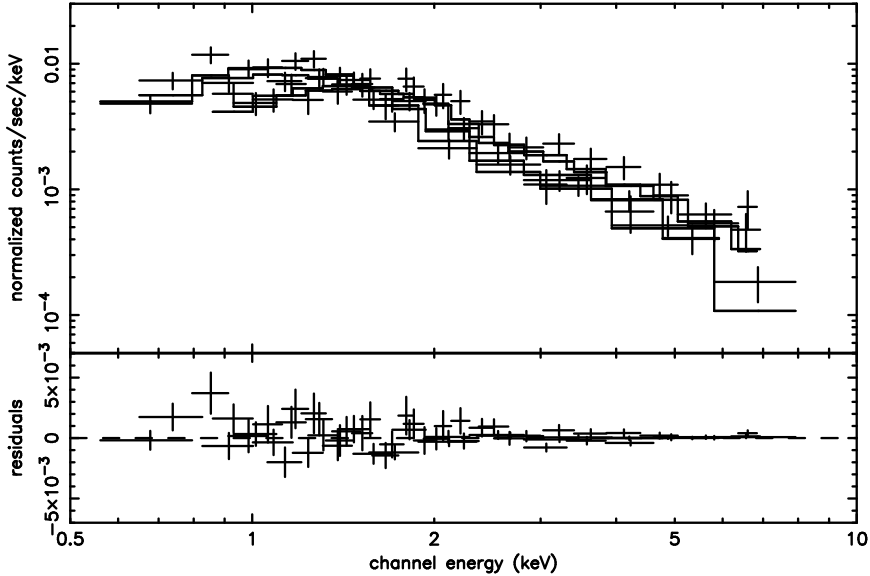}}
\caption{The best fit power-law fit to the \asca spectrum of  IRAS00317-2142
together with the corresponding residuals.} 
\end{figure*}

\section{DISCUSSION}

The detection of variability between the \asca and the \rosat data
 clearly suggests 
 an AGN origin for the X-ray emission. The \asca data are well
 represented with 
 a single power-law $\Gamma \sim 1.8$  with no absorption above the Galactic 
 $N_H\sim 1.5 \times 10^{20}$ $\rm cm^{-2}$. 
Hence, the X-ray spectrum alone suggests a Seyfert-1 type AGN. This
 is again compatible with the high X-ray luminosity of this object,
 $L_x\sim 10^{43}$ \lunits, during the \rosat observation. 
However, this interpretation comes in
 stark contrast with the optical 
spectrum which is indicative of a low  luminosity or obscured AGN.

Then a few possibilities arise for the nature of the AGN 
 in IRAS00317-2142. We  could view a Seyfert-1 nucleus 
 overwhelmed in the optical by the emission of
 a powerful starforming galaxy.
 In the X-ray band  the emission from the Seyfert-1 nucleus 
  should dominate over that arising from star-forming processes.  
 Still, according to the \asca spectral fits, the luminosity 
 of the RS component alone is $5\times 10^{41}$ $\rm erg~s^{-1}$;
 this classifies IRAS00317-2142 as one of the most 
 powerful X-ray star-forming galaxies known with 
 a luminosity more than an order of 
 magnitude above that of M82 (Ptak et al. 1997).  
  The above scenario for the composites, which was originally proposed by 
 Moran et al. (1996) can be tested by comparing the level 
 of the $H_{\alpha}$ 
 in respect with the hard X-ray luminosity.  
 Indeed, Ward et al. (1988) 
 found a strong correlation between the two quantities 
 in a sample of IRAS selected Seyfert-1 galaxies.
 Then according to the scenario above, the composites should follow
 the same relation between $L_x$ and broad $L(H_\alpha)$.   
The luminosity of the broad $H_{\alpha}$ component 
  in our object is about half of the total $H_\alpha$ luminosity
 (Moran et al. 1996). Then the observed broad $H_\alpha$ 
 luminosity is $L(H_\alpha) \sim  2\times  10^{41}$ \lunits 
 (Coziol et al. 1993). This roughly translates to an 
 X-ray luminosity of  $L_x\sim  4\times 10^{42}$ \lunits
   according to Ward et al. (1988) not far off the 
  observed X-ray luminosity of $L_x\sim  2\times 10^{42}$ \lunits. 
 Of course the long term X-ray variability  observed 
 introduces some level of uncertainty in the above test.
  Note that Bassani et al. (1999) reported the detection 
 of a few Seyfert galaxies which possibly have  a weak or absent 
 Broad Line Region. Our object  could in principle 
 belong to this category. However, the 
  $L_x/L(H_\alpha)$ ratio rather argues against this hypothesis.  
Again the possibility that the 
 X-ray emission remains relatively unabsorbed while the 
 optical suffers from additional obscuration 
 cannot be ruled out. 
 The Balmer decrement in our object (using the 
 flux of the narrow $H_\alpha$ and $H_\beta$ from Moran et al. 1996) 
 is about 5 corresponding to 
 a column of $N_H\sim 10^{21}$ $\rm cm^{-2}$ 
 (Bohlin et al. 1978) 
 ie an order of magnitude higher than the column 
 derived above in the case of cold absorption.
 The reason for this discrepancy is not obvious.   
 One possibility is that the absorbing column 
 is ionised: indeed  
 in the case of a warm absorber the derived column is 
 $N_H\sim 10^{22}$ $\rm cm^{-2}$. 
 Then some fraction of the 'warm' column should be located 
 further away from the nucleus, where the narrow 
 $H_\alpha$ and $H_\beta$ lines originate. 
 
 Finally, IRAS00317-2142 could 
 be a heavily obscured (Compton thick) AGN like eg NGC1068. 
 Then a large fraction of the X-ray emission  
 could be due to scattered X-rays, on a warm electron medium which should
 be situated well above the obscuring torus. 
 The broad $H_\alpha$ wing observed could arise from 
  scattered radiation. However, the 
 narrow and broad $H_\alpha$ components have comparable fluxes 
 (Moran et al. 1996) arguing against this interpretation. 
  Note that, the peculiar combination of
 an un-absorbed X-ray spectrum with a narrow-line dominated, 
 obscured optical spectrum was also  encountered in  
 NGC3147 (Ptak et al. 1996). 
 This object  has a type-2 type nucleus according to its optical 
 spectrum which presents a relatively broad [NII] line 
 (FWHM$\sim$400 $\rm km~s^{-1}$). The
 X-ray spectrum of NGC3147 is again very similar to our object as it
 shows no intrinsic 
absorption and a steep spectral index $\Gamma \approx 1.8$. Its
 observed X-ray luminosity is far below ($L_x\sim 10^{41}$ \lunits) 
 that of our object; an Fe line 
 at 6.4 keV (rest-frame) was clearly detected in NGC3147 (Ptak et
 al. 1996) with an 
 equivalent width greater than 130 eV. Ptak et al. (1996) favoured a
 scenario where NGC3147 harbours an obscured AGN.    
   If indeed IRAS00317-2142 is a heavily obscured AGN the intrinsic 
 unabsorbed luminosity
 of this object should exceed $10^{45}$ \lunits
 (assuming that we observe only a few percent of scattered 
 light alone our line of sight). 
 The column density should be above $N_H>10^{24}$ 
$\rm cm^{-2}$ in order to absorb all transmitted 
 photons with energy up to 7 keV.
 Such a large column should produce a high equivalent width Fe line as
 the  cold obscuring material is photoionised by the incident hard 
X-ray photons. For an obscuring column of $10^{24}$ $\rm cm^{-2}$
 we expect an 
 equivalent width Fe line of $\sim$ 1 keV (see Fig. 3 of Ptak et al. 1996). 
This value is consistent with our 90 per 
 cent upper limit for the Fe line equivalent
 width (0.9 keV). 
Probably the most stringent constraints 
 on the AGN nature come from the  
  X-ray to optical emission line ratios.
 Maiolino et al. (1998) argue that as the 
 [OIII] emission comes from scales much larger than the 
 obscuring torus, the 
 $f_x/f_{[OIII]}$ ratio  
 provides a powerful diagnostic 
  of the nuclear activity. 
 Hence, an obscured AGN should have very low 
  $f_x/f_{[OIII]}$ ratio as the nuclear X-ray luminosity
 is obscured while ethe [OIII] is not.
  The ratio of the X-ray (2-10 keV) 
 to the [OIII] flux (corrected for absorption
 using the formula of Bassani et al. 1999) 
  is about 2.5 for our object. 
 This is more typical of unobscured AGN 
 (Maiolino et al. 1998) 
  arguing against the obscured AGN scenario. 
    
Additional constraints on the nature of the AGN could in 
 principle, be provided 
 by the observed variability.
 The Seyfert-1 objects should present rapid variability
 with the amplitude increasing as we go to lower 
 luminosities (eg Nandra et al. 1997, Ptak et al. 1998). 
 In contrast, the obscured AGN show little or no variability 
 as a large fraction of the X-ray emission 
 comes from re-processed radiation far away from the 
 nucleus. However, in our case 
 the \asca light curves  have poor photon statistics.
 Both the GIS and the SIS data (4 ksec bins) 
 cannot rule out a constant 
 count rate even at the 68 per cent confidence level.
 Therefore it is difficult to differentiate, 
 on the basis of variability alone, between 
 a Seyfert-1 and an obscured AGN scenario.   

Regardless of the X-ray emission origin, 
the enigmatic composite objects present many
 similarities with the NLXGs detected in abundance in deep \rosat surveys 
(eg Boyle et al. 1995). Many of these present clear-cut star-forming galaxy 
spectra. However, their high X-ray luminosities immediately rule out a
 star-forming galaxy origin for the X-ray emission. 
Only \rosat data exist for these faint objects 
and therefore their X-ray spectra 
remain yet largely unconstrained (eg Almaini et al. 1996). 
Thus it is difficult to compare the X-ray properties of NLXGs 
with those of the 
'composites'. However, it is interesting to note that if the \rosat  
NLXGs have X-ray spectra similar to IRAS00317-2142, 
then they would make a small
 contribution to the X-ray background. Indeed, 
the X-ray spectrum of IRAS003217-2142 in the 2-10 keV band is 
much steeper than that of the X-ray background 
 in the same band (Gendreau et al. 1995).

\section{CONCLUSIONS}

The X-ray observations shed more light on the 
 nature of the composite objects. The detection 
 of strong variability 
 (a factor of three) between the \asca and the \rosat observations 
 clearly suggests an AGN origin for 
 most of the X-ray emission. 
  Some fraction of the 
 soft 0.1-2 keV X-ray emission can still be attributed 
 to a strong star-forming 
 component ($L_x\sim 10^{41-42} $ \lunits).  
 Nevertheless, the exact AGN classification remains uncertain. 
 The X-ray spectrum has a steep power-law slope ($\Gamma =1.8$)
 and presents no absorption above the Galactic. 
 Hence the X-ray spectrum is clearly suggestive of 
 a Seyfert-1 interpretation.  
 However, the optical spectrum shows only a weak $H_{\alpha}$ 
 component and is therefore more reminiscent of an obscured Seyfert
 galaxy. The discrepancy between the optical and the X-ray spectrum 
 can be alleviated if we assume that  
  while the optical spectrum is influenced by 
 the strong star-forming component, the 
 AGN component is producing most of the X-ray emission; 
 alternatively, it is possible that our object has a dusty 
 ionised absorber which selectively obscures the optical emission.  
 Finally, the possibility that our object is Compton thick 
 (possibly like eg NGC3147) is disfavoured 
 by the large value of the $f_x/f_{[OIII]}$ ratio 
 which is more typical of  unobscured AGN. 
  
Future imaging observations with {\it Chandra} and 
 high throughput spectroscopic observations with the {\it XMM} mission will 
provide the conclusive test on the above hypothesis.  

\section{Acknowledgments}
 I am grateful to the referee R. Maiolino for many useful 
 comments and suggestions. 
I would like to thank  A. Zezas for extracting the 
 \asca light curves. It is a pleasure to thank 
 A. Ptak, A. Zezas, I. Papadakis and G.C. Stewart for many useful 
 discussions. 
 This research has made use of data obtained through the High Energy 
Astrophysics Science Archive Research Center Online Service, provided
by the NASA/Goddard Space Flight Center.


\begin{thebibliography}{99}

\bibitem{} Almaini, O., Shanks, T., Boyle, B.J., Griffiths, R.E., 
  Roche, N., Stewart, G.C., Georgantopoulos, I., 1996, MNRAS, 282, 295 
\bibitem{} Bassani, L., Dadina, M., Maiolino, R., Salvati, M. 
 Risaliti, G., Della Ceca, R.,
 Matt, G., Zamorani, G. 1999, ApJS, 121, 473
\bibitem{}Bohlin,R.C., Savage, B.D., Drake, J.F., 1978, ApJ, 224, 132
\bibitem{}Boller, Th., Meurs, E.J.A., Brinkmann, W., Fink, H.,
 Zimmermann, U., Adorf, H.M.  1992, A\&A, 261, 57
\bibitem{}  Boller, Th., Bertoldi, F.,
 Dennefeld, M., Voges, W., 1998, A\&AS, 129, 87 
\bibitem{}Boyle, B.J., McMahon, R.G., Wilkes, B.J., Elvis, M., 1995, 
 MNRAS, 276, 315
\bibitem{} Brandt, W.N., Fabian, A.C., Pounds, K.A., 1996, 
 MNRAS, 278, 326
\bibitem{} Brandt, W.N., Gallagher, S.C., Laor, A., Wills, B.J., 1999,  
 Astroph. Letters \& Communications, in press  
\bibitem{} Coziol, R., Demers, S., Pena, M., Torres-Peimbert, S., 
 Fontaine, G., Wesemael, F., Lamontagne, R.,  1993, AJ, 105, 35 
\bibitem{} Ebeling, H., Voges, W., Boehringer, H., 1994, ApJ, 436, 44 
\bibitem{} Gendreau, K. et al. 1995, PASJ, 47, L5
\bibitem{} Iwasawa, K., Nandra, K., Fabian, A.C., 1999, MNRAS, 307, 611
\bibitem{} Maiolino, R., Salvati, M.,
 Bassani, L., Dadina, M., Della Ceca, R., Matt, G.,
 Risaliti, G., Zamorani, G., 1998, A\&A, 338, 781
\bibitem{}Moran, E.C., Halpern, J.P., Helfand, D.J.,  
 1996, ApJS, 106, 341 
\bibitem{} Nandra, K. \& Pounds, K.A.P., 1994, MNRAS, 268, 405 
\bibitem{} Ptak, A. Yaqoob, T., Serlemitsos, P.J., Kunieda, H., Terashima, Y., 
      1996, 459, 542
\bibitem{} Ptak, A., Serlemitsos, P., yaqoob, T., Mushotzky, R., 
 Tsuru, T.,  1997, AJ, 113, 1286
\bibitem{} Ptak, A., Yaqoob, T., Mushotzky, R., Serlemitsos, P., 
 Griffiths, R.E., 1998, ApJL, 501, L37 
\bibitem{}Pildis, R., Bregman, J.N.,  Evrard, A.E., 1995, ApJ, 443, 514
\bibitem{} Saracco, P. \& Ciliegi, P., 1995, A\& A, 301, 348 
\bibitem{} Tanaka, Y., Inoue, H., Holt, S.S., 1994, PASJ, 46, L37
\bibitem{} Veilleux, S. \& Osterbrock, D.E., 1987, ApJS, 63, 295 
\bibitem{} Voges, W. et al., 1996, IAUC, 6420, 2 
\bibitem{} Ward, M.J., Done, C., Fabian, A.C.,  Tennant, A.F., 
 Shafer, R.A., 1998, ApJ, 324, 767 
\bibitem{} Yaqoob, T., "The ASCA Data Reduction Guide v.2.0", GSFC/NASA, 1997
\end{thebibliography}
\end{document}